\documentclass[aps,prl,twocolumn, superscriptaddress]{revtex4-1}
\usepackage{graphicx,color}
\usepackage{amsmath,amssymb,latexsym}
\usepackage{mathtools}
\usepackage[english]{babel}
\usepackage{dcolumn}
\usepackage{bm}
\usepackage{float}
\usepackage[normalem]{ulem}

\usepackage{color}

\ProvideTextCommand{\DJ}{OT1}{\leavevmode\raisebox{-.5ex}{\makebox[0pt][l]{\hskip-.07em\accent"16\hss}}D}

\begin{document}

\title{Being heterogeneous is disadvantageous: Brownian non-Gaussian searches}

\author{Vittoria Sposini}
\email{vittoria.sposini@univie.ac.at}
\affiliation{
  Faculty of Physics, University of Vienna, Kolingasse 14-16, 1090 Vienna, Austria
}

\author{Sankaran Nampoothiri}
\email{sparampo@gitam.edu}
\affiliation{
Department of Physics, Gandhi Institute of Technology and Management (GITAM) University, Bengaluru-561203 , India
}

\author{Aleksei Chechkin}
\email{chechkin@uni-potsdam.de}
\affiliation{
  Faculty of Pure and Applied Mathematics, Hugo Steinhaus Center, Wroclaw University of Science and Technology, Wyspianskiego Str. 27, 50-370 Wroclaw, Poland
}
\affiliation{
  Institute for Physics \& Astronomy, University of Potsdam, 14476 Potsdam-Golm,
  Germany
}
\affiliation{
  Akhiezer Institute for Theoretical Physics, 61108 Kharkov, Ukraine
}

\author{Enzo Orlandini}
\email{enzo.orlandini@unipd.it}
\affiliation{
Dipartimento di Fisica e Astronomia `G. Galilei' - DFA, Sezione INFN,
Universit\`a di Padova,
Via Marzolo 8, 35131 Padova (PD), Italy
}

\author{Flavio Seno}
\email{flavio.seno@unipd.it}
\affiliation{
Dipartimento di Fisica e Astronomia `G. Galilei' - DFA, Sezione INFN,
Universit\`a di Padova,
Via Marzolo 8, 35131 Padova (PD), Italy
}

\author{Fulvio Baldovin}
\email{fulvio.baldovin@unipd.it}
\affiliation{
Dipartimento di Fisica e Astronomia `G. Galilei' - DFA, Sezione INFN,
Universit\`a di Padova,
Via Marzolo 8, 35131 Padova (PD), Italy
}

\date{\today}

\begin{abstract}
Diffusing diffusivity models, polymers in the grand canonical
ensemble and polydisperse, and  continuous time random walks,
all exhibit 
stages of non-Gaussian diffusion. Is non-Gaussian targeting more
efficient than Gaussian? We address this question, central to, e.g.,
diffusion-limited reactions and some biological processes, through a
general approach that makes use of Jensen's inequality and that
encompasses all these systems. In terms of customary
mean first passage time, we show that Gaussian searches are more
effective than non-Gaussian ones. A companion paper argues that
non-Gaussianity becomes instead highly more efficient 
in applications where only a small fraction of tracers is required
to reach the target.
\end{abstract}

\maketitle

\section{Introduction}
The Brownian non-Gaussian motion refers to the interesting contingency
of observing a stochastic process characterized by a mean squared
displacement which linearly increases in time -- Brownian or Fickian
behavior -- concomitant with a non-Gaussian probability density
function (PDF) for the displacements.  Since its discovery in a
variety of experimental
conditions~\cite{wang2009,wang2012,toyota2011,yu2013,yu2014,chakraborty2020,weeks2000,wagner2017,jeon2016,yamamoto2017,stylianidou2014,parry2014,munder2016,cherstvy2018,li2019,cuetos2018,hapca2008,pastore2021rapid}
and molecular dynamics
simulations~\cite{pastore2015,miotto2021length,pastore2022} it was
expected~\cite{wang2012} that the excess of probability for rare
fluctuations might dominate first-passage processes. Recent analyses
showed however that typical Gaussian searches turn out to be more
effective than non-Gaussian
ones~\cite{grebenkov2018,grebenkov2019,sposini2019first}.  This issue
finds here a general assessment encompassing different experimental
situations and theoretical models, including diffusing
diffusivities~\cite{chechkin2017}, polymers in the grand canonical
ensemble~\cite{nampoothiri2021,nampoothiri2022,marcone2022} and
polydisperse~\cite{flory1953,cosgrove2005,odian2004}, and continuous
time random
walks~\cite{klafter2011,barkai2020,wang2020,sokolov2021,sokolov2021ld}.
By using the Jensen's inequality~\cite{rudin1987} we first show that
the ``tail effect'' -- associated with faster diffusion -- is in fact
accompanied by a ``central effect'', i.e., an excess probability for
slower diffusion. The question then comes up about which one is
dominant in first-passage processes. The answer depends on the
threshold for the fraction of tracers reaching the target which is
relevant to the specific application.  A further implementation of
Jensen's inequality allows us to demonstrate that indeed the 
typical time scale for one searcher to reach the target, e.g. 
the mean first passage time is shorter in Gaussian than in non-Gaussian
diffusion. However, the scenario drastically changes if the relevant
physical time scale is instead related to the first few successful
searches among many: In this case, a companion
paper~\cite{sposini2023prl} shows that the non-Gaussian behavior
becomes significantly faster than the Gaussian one.

In the next Section, we introduce the general mechanism leading to
non-Gaussianity in subordination processes, highlighting the ``tail''
and ``central'' effects. After presenting various subordination
models, we then proceed to highlight two dynamical regimes,
characterized by different scaling properties of the PDF for the
subordinator. Paradigmatic targeting problems are then discussed
within this context, and we finally draw our conclusions.

\section{Faster and slower diffusion}
Let us first recall how Brownian non-Gaussian diffusion emerges in
subordination processes. Consider a situation in which some source of
heterogeneity makes the diffusion coefficient $D$ of overdamped
particles to fluctuate in time  (examples are provided below).
Indicating as $X(t)$ the random location along a certain axis
$x$ at time $t$ of the diffusing particle, we have
\begin{equation}
  \mathrm{d}X(t)
  =\sqrt{2\,D(t)}\,\mathrm{d}B(\mathrm{d}t)\,,
  \label{eq_lang_1}
 \end{equation}
where $B(t)$ is a Wiener process (Brownian motion), and $D(t)$ describes
the stochastic process associated with the fluctuating diffusion
coefficient.
We indicate as $p_D^*$ the steady-state distribution density of
$D(t)$, which could either be a PDF or a probability mass function
(PMF) depending on whether $D$ varies continuously or discretely,
and as $D_{\mathrm{av}}\equiv\mathbb{E}[D]$ its average value. 
Technically, it is convenient to introduce the \textit{subordinator}
process, defined as
\begin{equation}
  S(t)\equiv2\int_0^t\mathrm{d}t'\,D(t')
  \;\Rightarrow \mathrm{d}S=2\,D(t)\,\mathrm{dt}\,;
  \label{eq_subordinator_def}
\end{equation}
in such a way Eq.~\eqref{eq_lang_1} is reexpressed in the
\textit{random path} or \textit{subordinator
  parametrization}~\cite{chechkin2017,nampoothiri2021,nampoothiri2022,marcone2022}:
\begin{equation}
  \mathrm{d}X(t)=\mathrm{d}B(\mathrm{d}S)\,.
\end{equation}
The PDF for the tracer in position $x$
at time $t$, given that it was at $x_0$ at time zero is then obtained
through the subordination
formula~\cite{Feller1968,bochner2020harmonic} 
\begin{equation}
  \label{eq_subordination}
  p_X(x,t|x_0)=\int_0^\infty\mathrm{d}s\;G_{\mathrm{BG}}(x,s|x_0)\;p_S(s,t)\,,
\end{equation}
where $p_S(s,t)$ is the probability for the path parametrization $s$
at time $t$, and
$G_{\mathrm{BG}}(x,s|x_0)$ is the Green function for the
Brownian-Gaussian (BG) ordinary diffusion associated with the
problem's boundary conditions.

\begin{figure}[t]
  \includegraphics[width=0.95\columnwidth]{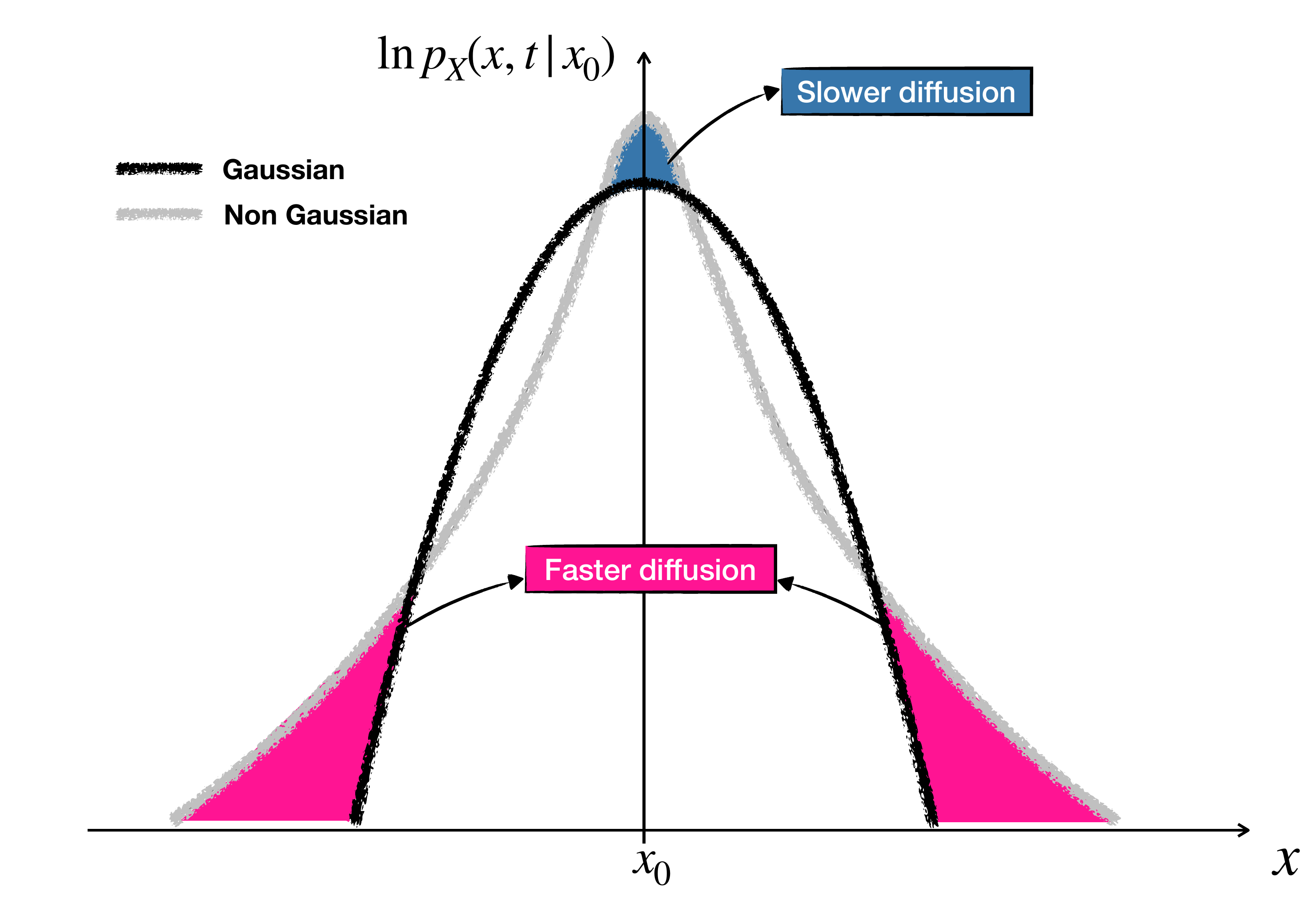}\\
  \caption{Comparison between Gaussian and non-Gaussian PDFs for
    subordination processes. The two PDFs share the same mean and standard
    deviation but the non-Gaussian one has an excess probability
    both in the tails and in the center part. 
    The non-Gaussian PDF is obtained from the FSP model with $p=0.99$ (See text).
    }
  \label{fig_excess_pdf}
\end{figure}

It is remarkable that, given the common subordination structure, quite
different stochastic models share the same qualitative
non-Gaussian features;
to introduce these
features,
let us first concentrate on free diffusion. 
In free diffusion (with natural boundary conditions),
\begin{equation}
  G_{\mathrm{BG}}(x,s|x_0)
  =\dfrac{\mathrm{e}^{-\frac{(x-x_0)^2}{2\,s}}}{\sqrt{2\pi\,s}}\,,
\end{equation}
and Eq.~\eqref{eq_subordination} already highlights the non-Gaussian nature of the diffusion, as the tracer's PDF is a
superposition of Gaussian PDFs.
A change of variable in
Eq.~\eqref{eq_subordination} shows how 
the moments of $X(t)$ are
linked to those of the subordinator:
\begin{equation}
  \mathbb{E}[(X(t)-x_0)^m]
  =G_{\mathrm{BG}}^{(m)}
  \int_0^\infty\mathrm{d}s\,p_S(s,t)\,s^{\frac{m}{2}}
  =G_{\mathrm{BG}}^{(m)}\,\mathbb{E}[S^{\frac{m}{2}}(t)],
  \label{eq_moments}
\end{equation}
where
\begin{equation}
  G_{\mathrm{BG}}^{(m)}\equiv
  \int_{-\infty}^{+\infty}\mathrm{d}x
  \,\dfrac{\mathrm{e}^{-\frac{(x-x_0)^2}{2}}}{\sqrt{2\pi}}
  \,(x-x_0)^{m}\,.
\end{equation}
In this paper, we focus on equilibrium initial conditions
for $D(t)$, i.e. we assume that $D(t)$ is distributed according to the
steady-state distribution $p_D^*$ so that
\begin{equation}
  \mathbb{E}[D(t)]=D_{\mathrm{av}}\,.
\end{equation}
Through Eq.~\eqref{eq_moments} with $m=2$ and
Eq.~\eqref{eq_subordinator_def}, this is sufficient to
guarantee the Brownian behavior:
\begin{equation}
  \mathbb{E}[(X(t)-x_0)^2]=\mathbb{E}[S(t)]
  =2\,D_{\mathrm{av}}\,t\,.
\end{equation} 
While Gaussian variables have zero excess kurtosis $\kappa_X(t)-3$, with the
kurtosis defined as
\begin{equation}
  \kappa_X(t)\equiv\dfrac{\mathbb{E}[(X(t)-x_0)^4]}{\left(\mathbb{E}[(X(t)-x_0)^2]\right)^2}\,,
\end{equation}
subordination processes are leptokurtic, that is, they are
characterized by a positive excess kurtosis.
This is again a consequence of Eq.~\eqref{eq_moments}: Since
$G_{\mathrm{BG}}^{(4)}/(G_{\mathrm{BG}}^{(2)})^2=3$, we have
\begin{equation}
  \kappa_X(t)-3
  =3
  \,\dfrac{
    \mathbb{E}[S^2(t)]-\left(\mathbb{E}[S(t)]\right)^2
  }{
    \left(\mathbb{E}[S(t)]\right)^2
  }
  >0\,.
  \label{eq_excess_kurtosis}
\end{equation}
As a result, subordination processes possess
an \textit{excess of probability in the tails} of the PDF,
compared to a Gaussian PDF with equal variance
(see Fig.~\ref{fig_excess_pdf}). This effect is
triggered by the faster diffusers in $p_D^*$.

On the other hand
the Jensen's inequality~\cite{rudin1987} says that for
a real-valued $\mu$-measurable function $f$ on a sample space $\Omega$
and a convex function $\varphi$
on the real numbers we have
\begin{equation}
  \int_\Omega\mathrm{d}\mu\;(\varphi\circ f)
  \geq\varphi\left(\int_\Omega\mathrm{d}\mu\;f\right)\,,
\end{equation}
where the ``$\circ$'' symbol means a composition
of the functions. The inequality becomes strict if $\varphi$ is strictly convex and the
measure $\mu$ is not induced by a constant random variable.
Note now that
$\varphi(s)\equiv G_{\mathrm{BG}}(x_0,s|x_0)=1/\sqrt{2\pi\,s}$ is a convex
function of $s$. Taking $\mathrm{d}\mu\equiv\mathrm{d}s\,p_S(s,t)$ and
$f(s)\equiv s$, we thus have that for all time $t$,
\begin{eqnarray}
  p_X(x_0,t|x_0)
 &=&\int_0^\infty\mathrm{d}s\,p_S(s,t)
  \,G_{\mathrm{BG}}(x_0,s|x_0)
  \nonumber\\
  &>&G_{\mathrm{BG}}\left(x_0,\int_0^\infty\mathrm{d}s\,p_S(s,t)\,s\bigg|x_0\right)
  \nonumber\\
  &=&G_{\mathrm{BG}}(x_0,\mathbb{E}[S(t)]|x_0)\,.
\end{eqnarray}
Since both $p_X(x,t|x_0)$ and $G_{\mathrm{BG}}(x,s|x_0)$ are continuous around $x_0$, there must exist a neighborhood of the center $x_0$ in which this inequality remains valid.
Thus, the non-Gaussian PDF also has an \textit{excess of probability
  in the central part} compared to the Gaussian PDF, due to slower tracers (again, please refer to
Fig.~\ref{fig_excess_pdf}).
The natural question to address is which of these effects is dominant
when considering targeting processes.

\section{Subordination stochastic models}
The subordination class includes a variety of
stochastic models, depending on the details of the subordinator:
\begin{itemize}
\item[(i)] Diffusing diffusivity (DD) models~\cite{chechkin2017} 
  are obtained assuming the diffusion coefficient to be
  the square of an Ornstein-Uhlenbeck process,
  $D(t)\equiv\boldsymbol{Y}^2(t)$, with
  \begin{equation}
    \mathrm{d}\boldsymbol{Y}(t)
    =-\dfrac{\boldsymbol{Y}(t)}{\tau} \, \mathrm{d}t
    +\sigma
    \,\mathrm{d}\boldsymbol{B}_{\boldsymbol{Y}}(\mathrm{d}t)\,.
  \end{equation}
  The dimension of the vector $\boldsymbol{Y}(t)$ is 
  $d_{\boldsymbol{Y}}\in\mathbb{N}^*$, $\tau$ is the 
  autocorrelation of the process, and $\sigma$ defines 
  the intensity of the fluctuations.  
  The steady-state PDF is given by
  \begin{equation}
    p_D^*(D)
    =\dfrac{
      D^{d_{\boldsymbol{Y}}/2-1}\;\mathrm{e}^
      {-d_{\boldsymbol{Y}}\,D/(2\,D_{\mathrm{av}})}
    }{
      (2\,D_{\mathrm{av}}/d_{\boldsymbol{Y}})^{d_{\boldsymbol{Y}}/2}
      \;\Gamma(d_{\boldsymbol{Y}}/2)
    }\,,
  \end{equation}
  where $D_{\mathrm{av}}=\sigma^2 \tau d_{\boldsymbol{Y}}/2$ is the 
  average diffusion coefficient. 
  Under the name of ``stochastic volatility'',
  these models are used in finance to correct the Black-Scholes option
  pricing for non-Gaussian effects~\cite{heston1993,sircar2000}.

  Simulation of the DD model can be simply realized by updating in
  parallel two Ornstein-Uhlenbeck processes: The one for
  $\boldsymbol{Y}(t)$ and the one for $X(t)$. In the latter, at each
  update the increment is drawn from a normal distribution with zero
  average and variance $2\,D(t)=2\,\boldsymbol{Y}^2(t)$. The
  simulation time can be expressed in terms of $\tau$ and the other 
  two free parameters are $\sigma$ and $d_{\boldsymbol{Y}}$. 
  
\item[(ii)] A concrete simple example of a subordination process is
  offered by a polymer in a diluted solution, exchanging monomers with
  a chemostat: the Grand canonical polymer (GCP)
  model~\cite{nampoothiri2021,nampoothiri2022,marcone2022}.   
  Indeed, the center of mass of a polymer in solution is
  known to diffuse with a coefficient $D$ which depends on the number
  of monomers $N$ as $D(N)=D_1/N^\alpha$~\cite{deGennes1979,Doi1992}, with $D_1$ the diffusion coefficient of a single monomer.
  The value of $\alpha$ depends on the specific polymer model; for
  definiteness in this paper we adopt the Rouse value $\alpha=1$, but
  similar results apply to other models, such as the Zimm or the
  reptation ones~\cite{Doi1992}.  In the grand canonical
  ensemble $N$
  fluctuates in time, becoming a second source of noise besides the
  solvent collisions responsible for the Brownian motion.
  $N(t)$ can be
  simply modeled in terms of a birth-death
  process;
  in the mean-field limit, both the
  birth $\lambda$ and death $\mu$ reaction rates are independent of the polymer
  size and their ratio $p\equiv\lambda/\mu$
  corresponds to the ratio between the fugacity
  of the system and the critical fugacity~\cite{nampoothiri2022}:
  As $p\to1^-$, the average polymer size becomes infinite and
  relative size fluctuations diverge.
  The steady-state size distribution is
  \begin{equation}
    p_N^*(n)=(1-p)\,p^{n-1}
    \quad\textrm{for $n=1,2,\ldots$}\,,
  \label{eq_pstar_n_gcp}
  \end{equation}
  corresponding to the diffusion coefficient PMF
  \begin{equation}
  p_D^*(D_n)
  =\left(1-p\right)\,p^{D_1/D_n-1}\,,
  \label{eq_pstar_d_gcp}
  \end{equation}
  with $D_n=D_1/n$.

  Also for the GCP model simulations are realized through a parallel
  update, in this case of the processes $N(t)$ and $X(t)$. A simple
  way to simulate the birth-death process $N(t)$ is by implementing
  the Gillespie algorithm~\cite{gillespie1977} with reaction rates
  $\lambda$, $\mu$.
  As reported for instance in Ref.~\cite{nampoothiri2022},
  it is possible to approximate the
  autocorrelation time $\tau$ of $N(t)$ as
  \begin{equation}
    \tau=\dfrac{1+p}{(1-p)^2\;\mu}\,,
    \label{eq_tau_gcp}
  \end{equation}
  where $p=\lambda/\mu$.
  It is clear from
  Eq.~\eqref{eq_tau_gcp} that $\tau$ diverges as $p\to1^-$, a phenomenon
  called \emph{critical slowing down}.
  While the ratio $p=\lambda/\mu$ fixes the
  steady-state distribution $p_N^*$ and how close the simulation is to
  critical conditions, the parameter $\mu$ can still independently
  be fixed to calibrate the simulation time in terms of $\tau$,
  according to Eq.~\eqref{eq_tau_gcp}.  The remaining free parameter
  is the single-monomer diffusion coefficient $D_1$.
\item[(iii)] After polymerization terminates in a step-growth
  polymerization~\cite{odian2004}, one is left with a polydisperse sample of
  polymers with heterogeneous size $N$. Assuming chains with one 
  reaction center in the end, the size distribution coincides with
  Eq.~\eqref{eq_pstar_n_gcp} and it is called in this context
  Flory-Schulz distribution~\cite{odian2004}, with $p$ the
  \textit{reaction extent}. In this case, $D$ must be regarded as 
  a static random variable, $D(t)=D\;\forall t$, distributed according 
  to Eq.~\eqref{eq_pstar_d_gcp}.

  To simulate the behavior of the Flory-Schulz  polydisperse (FSP) model, 
  for each realization one simply picks a value of $D_n$ with  probability 
  $p_D^*(D_n)$, and simulate then the Langevin process $X(t)$ keeping 
  the diffusion coefficient fixed. 
  The statistic of the process is then obtained by averaging over
  the different histories. Free parameters are $p$ and $D_1$.
\item[(iv)] In the continuous time random walk
  (CTRW)~\cite{klafter2011,barkai2020,wang2020,sokolov2021ld} with
  average waiting time $\tau<\infty$, one starts with a discrete
  subordinator $K(t)$, with $p_K(k,t)$ the PMF for $K=k\in\mathbb{N}$
  steps operated at time $t$. If $X(k)$ is a simple random walk
  providing the location after $k$ steps with PDF $p_{X_k}(x,k|x_0)$,
  then one gets a discrete analogous of Eq.~\eqref{eq_subordination}:
  \begin{equation}
    p_X(x,t|x_0)=\sum_{k=0}^\infty p_{X_k}(x,k|x_0)\;p_K(k,t)\,.
    \label{eq_subordination_ctrw}
  \end{equation}
  For time $t\gg\tau$ the typical number of steps is very large, and
  the operational time $S(t)\substack{\equiv\\t\gg\tau}K(t)$ can be
  taken to be continuous, so that $p_S(s,t)$ represents the
  probability for the continuous number of steps $s$ operated at time
  $t$.  Correspondingly, a typical random walk $X(k)$ with finite
  variance tends to the Gaussian limit, recovering the subordination
  equation in the continuous form, Eq.~\eqref{eq_subordination}.

  Simulation of a CTRW simply proceeds as an ordinary random walk,
  once the value of the elapsed time taken by the update step is drawn
  from the assumed waiting-time distribution. Free parameters in the
  simulations are those defining the waiting-time distribution, in
  particular its average value $\tau$, and the length $L$ of the
  random-walk step. 
\end{itemize}
All these examples share the qualitative non-Gaussian features
reported in Fig.~\ref{fig_excess_pdf}.

\section{Asymptotic regimes}
\begin{figure}[t]
  \includegraphics[width=0.99\columnwidth]{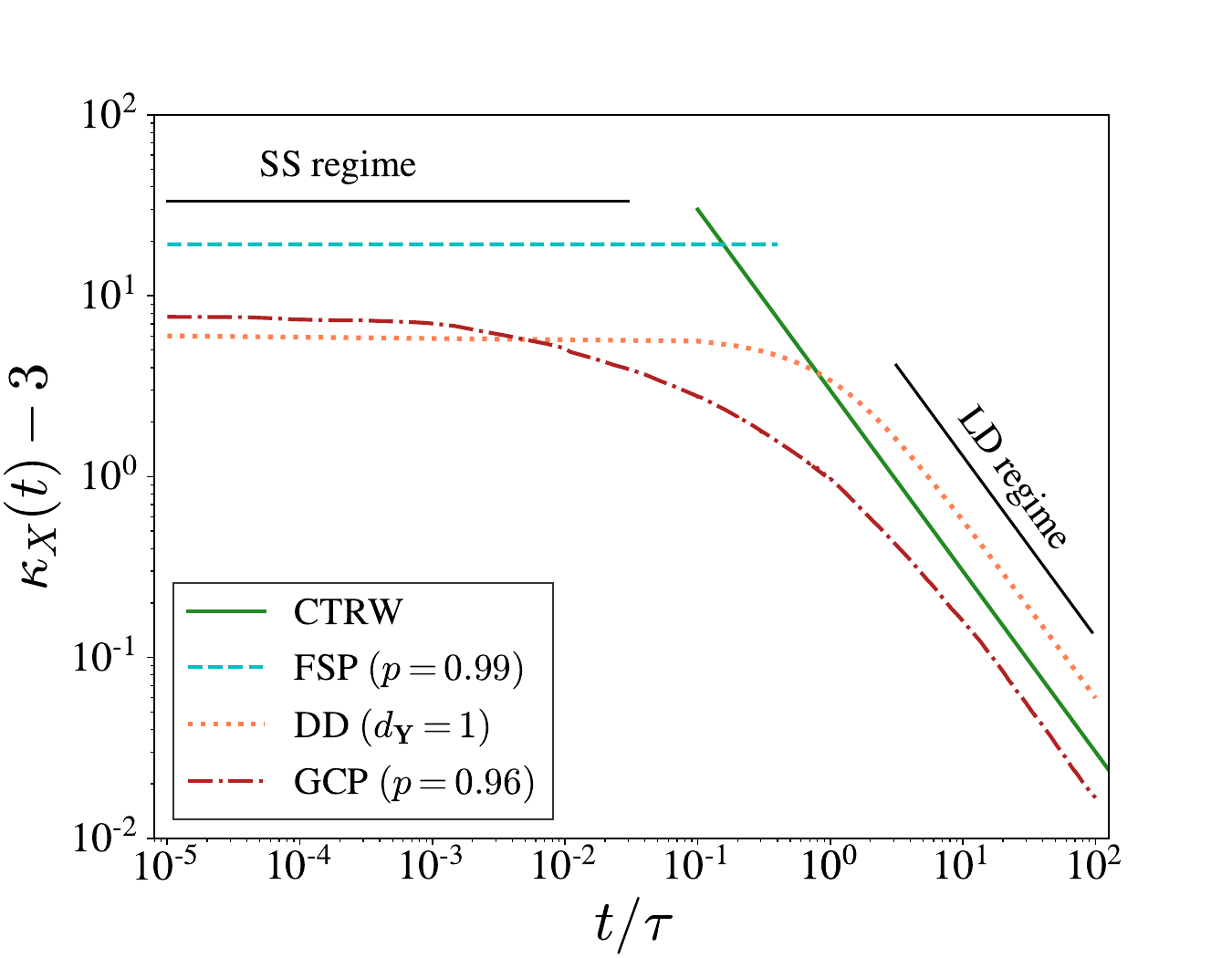}\\
  \caption{Time dependence of the excess kurtosis. A constant, positive
    $\kappa_X-3$ characterizes the SS regime, whereas in the LD regime 
    $\kappa_X(t)-3\sim1/t$  (black solid lines).
    While for the GCP, DD, and CTRW models the excess kurtosis 
    is  reported as a function of the rescaled
    time  $t/\tau$, for the FSP model it is plotted vs $t$.
    With these arrangements, the excess kurtosis depends only on the
    parameters displayed in the legend.  
    Curves for the DD and the GCP models have been obtained simulating
    the models as described in the text; those for the FSP model and the CTRW 
    with exponential waiting-time distribution have been exactly calculated. 
  }
  \label{fig_fixed_points}
\end{figure}

Before addressing search processes, some details about the dynamics
are needed. 
Subordination processes
may display two different regimes during which the PDF $p_S(s,t)$ can
be approximated in different ways. According to
Eq.~\eqref{eq_excess_kurtosis}, the excess kurtosis evolves
differently during the two regimes. 
\begin{itemize}
\item \textit{Super-statistics (SS) regime}.
  Consider a situation in which the diffusion coefficient is almost
  static, $D(t)\simeq D\;\forall t$,
  distributed according to $p_D^*$.
  This approximation is exact for the FSP model, case (iii) above, but it is also valid for cases (i) and (ii) taking a heterogeneous sample of tracers initially characterized by 
  the distribution $p_D^*$, as long as we consider time $t\ll\tau$.
  Indeed, for times much smaller than the autocorrelation time each
  tracer in the DD and GCP models basically retains its initial 
  diffusion coefficient.
  Within this approximation we have
  \begin{equation}
    S(t)=2\,D\,t\,,
  \end{equation}
  and a change of variable yields
  \begin{equation}
    p_S(s,t)=\frac{1}{2t}\;p_D^*\left(\frac{s}{2t}\right)\,.
    \label{eq_anomalous_scaling_1}
  \end{equation}
  From Eq.~\eqref{eq_excess_kurtosis}, the excess kurtosis is 
  \begin{eqnarray}
    \kappa_X(t)-3
    &=&3
    \,\dfrac{
      \mathbb{E}[D^2]-\left(\mathbb{E}[D]\right)^2
  }{
      \left(\mathbb{E}[D]\right)^2
    }
    \nonumber\\
    &=&const.>0
    \quad\textrm{(SS regime)}\,.
  \end{eqnarray}
  The SS regime is thus non-Gaussian;
  the behavior of the tracers can
  be characterized by operating an average of
  $D$-dependent quantities, over $p_D^*$.
  This ``superposition of statistics'' has been named in the
  literature
  super-statistics~\cite{beck2003,beck2006,hapca2008,wang2012},
  explaining the origin of the name.
\item \textit{Large deviation (LD) regime}.
  With $t\gg\tau$ the subordinators of 
  the DD and GCP models inherit
  from the Markovian evolution of $D(t)$ a LD
  principle~\cite{touchette2009}: 
  \begin{equation}
    p_S(s,t)\asymp\mathrm{e}^{-t\,I_S(s/t)}\,.
    \label{eq_ld}
  \end{equation}
  Here, ``$\asymp$'' stands
  for ``the dominant part as \mbox{$t\to\infty$}'', and $I_S$ is a rate
  function.
  Eq.~\eqref{eq_ld} implies a time
  evolution of the kurtosis $\kappa_X(t)$ different from the previous
  one. 
  The \textit{cumulant generating function} of $S(t)$ is defined as
  \begin{eqnarray}
    K_{S}(k,t)&\equiv&\ln\mathbb{E}\left[\mathrm{e}^{k\,S(t)}\right]
    \nonumber\\
    &=&\sum_{m=1}^\infty\dfrac{K_{S}^{(m)}(t)}{m!}\,k^m\,,
  \end{eqnarray}
  where $K_{S}^{(m)}(t)$ is the cumulant of order $m$ of $S(t)$.
  Using Eq.~\eqref{eq_ld} and the Laplace method one has
  \begin{equation}
    K_{S}(k,t)=t\,\lambda_{S/t}(k)\,,
  \end{equation}
  with the \textit{scaled} cumulant generating function~\cite{touchette2009} of 
  $S(t)/t$, $\lambda_{S/t}$, being time-independent. This means that
  within the LD approximation all the cumulants of $S(t)$ scale
  linearly with time, \mbox{$K_{S}^{(m)}(t)\propto t$}.
  As a consequence, from
  Eq.~\eqref{eq_excess_kurtosis} we obtain
  \begin{equation}
    \kappa_X(t)-3
    =\dfrac{K_{S}^{(2)}(t)}{\left(K_{S}^{(1)}(t)\right)^2}
      \propto1/t
    \;\;\textrm{(LD regime)}\,.
    \label{eq_excess_kurtosis_ld}
  \end{equation}
  Within this regime, the central part of $p_X(x,t)$ becomes Gaussian
  (central limit theorem~\cite{touchette2009}).
  As time passes by,
  the non-Gaussian behavior is relegated to larger and larger
  (lesser and lesser probable) fluctuations.
  The probability of the scaled subordinator
  $S(t)/t$ concentrates around its average value, and
  (apart from large deviations) the typical behavior is a 
  BG diffusion with coefficient $D_{\mathrm{av}}$.
  Besides characterizing   the DD and GCP models as
  $t\gg\tau$~\cite{chechkin2017,nampoothiri2021,nampoothiri2022,marcone2022},
  one can directly calculate that the CTRW, with an exponential waiting
  time distribution satisfies
  a LD principle~\cite{sokolov2021ld} and
  Eq.~\eqref{eq_excess_kurtosis_ld} 
  (namely, $\kappa_X(t)-3=3\,\tau/t$), for all time $t\geq0$.
\end{itemize}
Fig.~\ref{fig_fixed_points} displays the time evolution of the
excess kurtosis for the
different models, highlighting the two regimes. 

\section{Non-Gaussian and Gaussian targeting}
Let us now consider two classic targeting problems in one dimension:
\begin{itemize}
\item[a)] Finite interval $[0,L]$ with absorbing boundaries at the
  extrema. 
\item[b)] Semi-infinite domain $[0,\infty[$ with absorbing boundary at
    $x=0$.
\end{itemize}
For ordinary diffusion, exact expressions~\cite{redner2001}
are available for
the (cumulative) probability  of reaching a target by time $t$,
given the initial position $x_0$ and the diffusion coefficient $D$,
\begin{equation}
  P_T(t|x_0, D)=1-\mathcal{S}_T(t|x_0, D)
  =1-\int\mathrm{d}x\,G_{\mathrm{BG}}(x,t|x_0,D)
\end{equation}
($\mathcal{S}_T$ is the survival probability):
\begin{itemize}
\item[a)] For the finite interval,
  \begin{eqnarray}
    G_{\mathrm{BG}}(x,t|x_0,D)
    &=&\dfrac{2}{L}\sum_{k=1}^\infty
    \sin\left(\dfrac{k\,\pi}{L}\,x\right)
    \;\sin\left(\dfrac{k\,\pi}{L}\,x_0\right)\cdot
    \nonumber\\
    &&\qquad\cdot\;\mathrm{e}^{-\left(\frac{k\,\pi}{L}\right)^2\,D\,t}\,,
  \end{eqnarray}
  implying
  \begin{equation}
    P_T(t|x_0, D)
    =1-\dfrac{4}{\pi}\sum_{k=0}^\infty\sin\left(\dfrac{k\,\pi}{L}\,x_0\right)
    \dfrac{\mathrm{e}^{-\left(\frac{k\,\pi}{L}\right)^2\,D\,t}}{2k+1}\,.
  \end{equation}
\item[b)] For the semi-infinite domain,
  \begin{equation}
    G_{\mathrm{BG}}(x,t|x_0,D)
    =\dfrac{1}{\sqrt{4\pi,D\,t}}
    \left[
      \mathrm{e}^{-\frac{(x-x_0)^2}{4\,D\,t}}
      -\mathrm{e}^{-\frac{(x+x_0)^2}{4\,D\,t}}
      \right]\,,
  \end{equation}
  yielding
  \begin{equation}
    P_T(t|x_0, D)
    =1-\mathrm{erf}\left(\dfrac{x_0}{\sqrt{4\,D\,t}}\right)\,.
  \end{equation}
\end{itemize}
The time derivative of these expressions provides the PDF for reaching the
boundary at time $t$,  $p_T(t|x_0, D)=\partial_tP_T(t|x_0,D)$, from
which one can obtain the characteristic time for a single particle
to reach the target, $\tau_T$:
\begin{itemize}
\item[a)] With the finite domain, $\tau_T$ is naturally
  given by the mean
  first passage time,
  \begin{eqnarray}
    \tau_T(x_0,D)&=&\int_0^\infty\mathrm{d}t\,t\,p_T(t|x_0, D)
    \nonumber\\
    &=&\dfrac{x_0\,(L-x_0)}{2\,D}\,.
  \end{eqnarray}
\item[b)] The mean first passage time of the semi-infinite domain is infinite; however, a characteristic 
  time can still be identified as~\cite{redner2001} 
  \begin{eqnarray}
    \tau_T(x_0,D)
    &=&\left[\int_0^\infty\mathrm{d}t\,t^\beta\,p_T(t|x_0,D)\right]^{1/\beta}
    \nonumber\\
    &=&\dfrac{[\Gamma(1/2-\beta)]^{1/\beta}}{4\,\pi^{1/(2\beta)}}\dfrac{x_0^2}{D}\,,
  \end{eqnarray}
  for $\beta<1/2$.
\end{itemize}

\begin{figure}[t]
  \includegraphics[width=0.99\columnwidth]{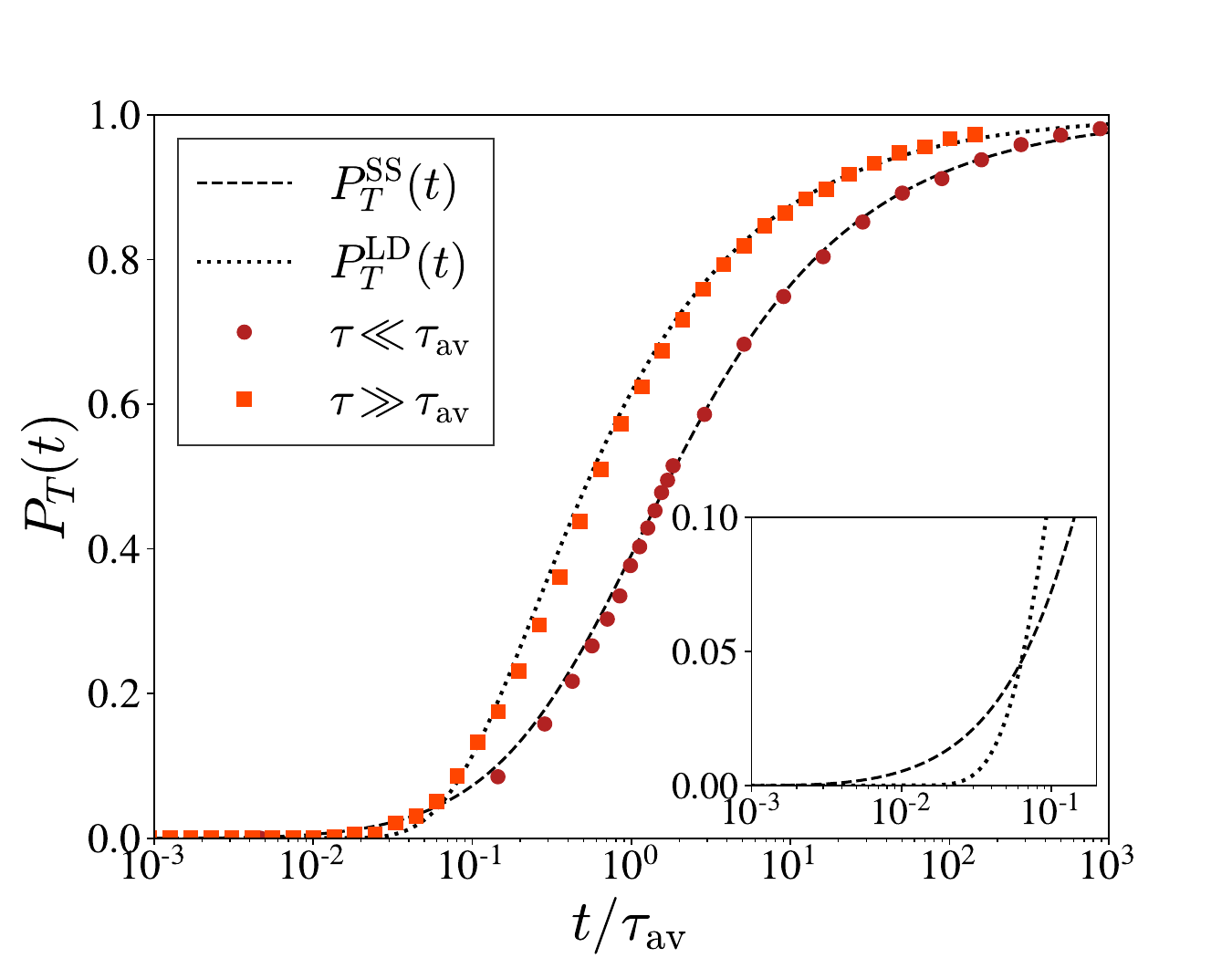}\\
  \caption{
    Gaussian and non-Gaussian targeting for the GCP at
    $p=0.99$. Dashed and dotted lines reproduce
    $P_T^{\mathrm{SS}}(t)$ and $P_T^{\mathrm{LD}}(t)$,
    respectively. The time axis is rescaled by $\tau_{\mathrm{av}}$
    defined in Eq.~\eqref{eq_tau_av} so that the plot is independent
    of $x_0$. Colored filled symbols are obtained simulating the birth-death
    polymerization process and the consequent polymer diffusion with
    the appropriate, time-varying diffusion coefficient; values for $\mu$ in
    Eq.~\eqref{eq_tau_gcp} has been chosen to satisfy the limits
    reported in the legend.
    Inset: magnification of $P_T^{\mathrm{SS}}(t)$ and
    $P_T^{\mathrm{LD}}(t)$ for small time.
  }
  \label{fig_probability}
\end{figure}

Given the two dynamical regimes discussed above, it is meaningful
to contemplate the probability associated with the SS regime, 
\begin{eqnarray}
  P_T^{\mathrm{SS}}(t|x_0)
  &\equiv&
  \int_0^\infty\mathrm{d}D\,p_D^*(D)\,P_T(t|x_0,D)
  \\
  &\equiv&
  \sum_{n=0}^\infty p_D^*(D_n)\,P_T(t|x_0,D_n)\,,
\end{eqnarray}
and the one corresponding to the LD regime,
\begin{equation}
  P_T^{\mathrm{LD}}(t|x_0)\equiv P_T(t|x_0,D_{\mathrm{av}})\,.
\end{equation}
According to Fig.~\ref{fig_fixed_points} the behavior of the FSP model
is characterized by the probability $P_T^{\mathrm{SS}}$
while the CTRWs for $t\gg\tau$ are ruled by $P_T^{\mathrm{LD}}$.  For the
DD model and the GCPs the actual probability will be close to
$P_T^{\mathrm{SS}}$ for 
$t\ll\tau$ and will tend to $P_T^{\mathrm{LD}}$ as $t\gg\tau$.
In Fig.~\ref{fig_probability} we highlight the two regimes for the GCP
model with the semi-infinite domain; a similar plot is valid for the
DD model~\cite{sposini2019first}. The abscissa is rescaled by the characteristic time for the particle with an average diffusion coefficient to travel over the
distance from the target, namely
\begin{equation}
  \tau_{\mathrm{av}}\equiv\dfrac{x_0^2}{2\;D_{\mathrm{av}}}\,.
  \label{eq_tau_av}
\end{equation}

As shown above, $\tau_T(x_0,D)$ is a convex function of $D$.
Taking now $\varphi(D)\equiv\tau_T(x_0,D)$, and again $f(D)=D$,
the Jensen's inequality\cite{rudin1987} conveys the following
basic result:
\begin{equation}
  \tau_T^{\mathrm{SS}}(x_0)\equiv\mathbb{E}[\tau_T(x_0,D)]
  >\tau_T^{\mathrm{LD}}(x_0)\equiv\tau_T(x_0,D_{\mathrm{av}})\,.
\label{eq_tau}
\end{equation}
Since Jensen's inequality is valid when averaging on general
distributions, Eq.~\eqref{eq_tau} applies to the DD and the GPC models,
and also for CTRWs if we compare the
early behavior ruled by the discrete subordination formula,
Eq.~\eqref{eq_subordination_ctrw}, with the LD Gaussian limit, attained
for $t\gg\tau$.  We thus conclude, consistently with earlier specific
findings~\cite{grebenkov2018,grebenkov2019,sposini2019first}, that
\textit{for a single searcher within the general class of non-Gaussian
  diffusion processes based on subordination, the characteristic time
  to target is larger than in ordinary diffusion}.  As the
characteristic time to target in ordinary diffusion
is in general inversely related to $D$, this
finding extends to general transient regimes and targeting processes,
not necessarily one-dimensional.
It is an effect due to the excess of
probability in the central part of the PDF (slower diffusers) (see
Fig.~\ref{fig_excess_pdf}) and it is somehow at odds with the general prospect of surprising phenomena triggered by rare fluctuations~\cite{wang2012}.

What about the ``tail effect'' in Fig.~\ref{fig_excess_pdf}?
By inspecting $P_T^{\mathrm{SS}}(t|x_0)$ and
$P_T^{\mathrm{LD}}(t|x_0)$, one finds that typically the latter is
larger than the former, consistently with Eq.~\eqref{eq_tau}. 
However, a closer look reveals that the ``tail effect'' dominates
at time shorter than $\tau_T^*$  (Fig.~\ref{fig_probability}, inset), 
with $\tau_T^*$ being the solution of
\begin{equation}
  \sum_{n=1}^\infty p_D^*(D_n)\,P_T(\tau_T^*|x_0,D_n)
  =P_T(\tau_T^*|x_0,D_{\mathrm{av}})\,.
\end{equation}
The corresponding fraction $P_T(\tau_T^*|x_0,D_{\mathrm{av}})$
of successful single-particle searches below which non-Gaussian chases
are more efficient than  the BG ones
is typically small; for instance
$\tau_T^*\simeq6.3\times10^{-2}\tau_{\mathrm{av}}$ and
$P_T(\tau_T^*|x_0,D_{\mathrm{av}})\simeq4.6\times10^{-2}$ in
Fig.~\ref{fig_probability}.
However, this apparently negligible effect makes a drastic difference
in \textit{extreme} first passage problems, where a small fraction of
the total number of searchers is required to first reach the target to
activate a certain function. This finding is discussed in
details in the
companion paper, Ref.~\cite{sposini2023prl}.

\section{Conclusions}
Targeting of receptors by ligand particles is a fundamental biological mechanism used by cells to activate or stop specific
functions. The mechanism makes use of the diffusive properties of the
ligands and the basic quantity to measure is the characteristic time employed by
a searcher to reach the target, i.e., when it exists finite, the mean
first passage time.
Since recent experiments and simulations highlighted heterogeneous conditions in which diffusion becomes Brownian non-Gaussian, the question arose whether non-Gaussianity enhances or not searches.

While this issue has already been addressed for specific
models~\cite{grebenkov2018,grebenkov2019,sposini2019first},
here we have provided a general approach valid for all
Brownian non-Gaussian processes based on subordination. 
Using Jensen's inequality and the positiveness of the variance of
the subordinator, we have demonstrated that 
the distributions of subordinated diffusive particles display an
excess of probability both in the central part and in the tails, when
compared with Gaussians.
The qualitative features appearing in Fig.~\ref{fig_excess_pdf} are independent of the kind of subordinator. At variance, the dynamical regimes shown in Fig.~\ref{fig_fixed_points} depend on the specific definition of the subordinator.  The DD and GCP models display both the SS and the LD regimes. The FSP model is characterized by the SS regime only, and the CTRW model with exponential waiting times possesses exclusively the LD regime. 

In agreement with earlier findings, we have shown on general ground
that the ``central effect'' dominates the one-particle searching process, making the characteristic time to target larger than in ordinary Gaussian
diffusion. This conclusion encompasses current models of Brownian
non-Gaussian diffusion and implies that, for ordinary diffusion-limited-reaction scenarios, non-Gaussianity weakens reaction rates.

The ``tail effect'' pertains to the realm of rare events and control
instead, \textit{extreme} searches, where only a few among many diffusers
are required to reach the target. In
Ref.~\cite{sposini2023prl} it is shown that this is the context in which
non-Gaussianity makes a substantial difference.

\section*{Acknowledgments}
F.S. and F.B. acknowledge the support by the project MUR-PRIN 2022ETXBEY, "Fickian non-Gaussian diffusion in static and dynamic
environments", funded by the European Union – Next Generation EU"'. V.S. acknowledges the support from the 
European Commission through the Marie Sk\l{}odowska-Curie COFUND 
project REWIRE, grant agreement No. 847693. A.C. acknowledges the support of the Polish National Agency for Academic Exchange (NAWA).


\begin{thebibliography}{49}%
\makeatletter
\providecommand \@ifxundefined [1]{%
 \@ifx{#1\undefined}
}%
\providecommand \@ifnum [1]{%
 \ifnum #1\expandafter \@firstoftwo
 \else \expandafter \@secondoftwo
 \fi
}%
\providecommand \@ifx [1]{%
 \ifx #1\expandafter \@firstoftwo
 \else \expandafter \@secondoftwo
 \fi
}%
\providecommand \natexlab [1]{#1}%
\providecommand \enquote  [1]{``#1''}%
\providecommand \bibnamefont  [1]{#1}%
\providecommand \bibfnamefont [1]{#1}%
\providecommand \citenamefont [1]{#1}%
\providecommand \href@noop [0]{\@secondoftwo}%
\providecommand \href [0]{\begingroup \@sanitize@url \@href}%
\providecommand \@href[1]{\@@startlink{#1}\@@href}%
\providecommand \@@href[1]{\endgroup#1\@@endlink}%
\providecommand \@sanitize@url [0]{\catcode `\\12\catcode `\$12\catcode
  `\&12\catcode `\#12\catcode `\^12\catcode `\_12\catcode `\%12\relax}%
\providecommand \@@startlink[1]{}%
\providecommand \@@endlink[0]{}%
\providecommand \url  [0]{\begingroup\@sanitize@url \@url }%
\providecommand \@url [1]{\endgroup\@href {#1}{\urlprefix }}%
\providecommand \urlprefix  [0]{URL }%
\providecommand \Eprint [0]{\href }%
\providecommand \doibase [0]{http://dx.doi.org/}%
\providecommand \selectlanguage [0]{\@gobble}%
\providecommand \bibinfo  [0]{\@secondoftwo}%
\providecommand \bibfield  [0]{\@secondoftwo}%
\providecommand \translation [1]{[#1]}%
\providecommand \BibitemOpen [0]{}%
\providecommand \bibitemStop [0]{}%
\providecommand \bibitemNoStop [0]{.\EOS\space}%
\providecommand \EOS [0]{\spacefactor3000\relax}%
\providecommand \BibitemShut  [1]{\csname bibitem#1\endcsname}%
\let\auto@bib@innerbib\@empty
\bibitem [{\citenamefont {Wang}\ \emph {et~al.}(2009)\citenamefont {Wang},
  \citenamefont {Anthony}, \citenamefont {Bae},\ and\ \citenamefont
  {Granick}}]{wang2009}%
  \BibitemOpen
  \bibfield  {author} {\bibinfo {author} {\bibfnamefont {B.}~\bibnamefont
  {Wang}}, \bibinfo {author} {\bibfnamefont {S.~M.}\ \bibnamefont {Anthony}},
  \bibinfo {author} {\bibfnamefont {S.~C.}\ \bibnamefont {Bae}}, \ and\
  \bibinfo {author} {\bibfnamefont {S.}~\bibnamefont {Granick}},\ }\href@noop
  {} {\bibfield  {journal} {\bibinfo  {journal} {Proceedings of the National
  Academy of Sciences}\ }\textbf {\bibinfo {volume} {106}},\ \bibinfo {pages}
  {15160} (\bibinfo {year} {2009})}\BibitemShut {NoStop}%
\bibitem [{\citenamefont {Wang}\ \emph {et~al.}(2012)\citenamefont {Wang},
  \citenamefont {Kuo}, \citenamefont {Bae},\ and\ \citenamefont
  {Granick}}]{wang2012}%
  \BibitemOpen
  \bibfield  {author} {\bibinfo {author} {\bibfnamefont {B.}~\bibnamefont
  {Wang}}, \bibinfo {author} {\bibfnamefont {J.}~\bibnamefont {Kuo}}, \bibinfo
  {author} {\bibfnamefont {S.~C.}\ \bibnamefont {Bae}}, \ and\ \bibinfo
  {author} {\bibfnamefont {S.}~\bibnamefont {Granick}},\ }\href@noop {}
  {\bibfield  {journal} {\bibinfo  {journal} {Nature materials}\ }\textbf
  {\bibinfo {volume} {11}},\ \bibinfo {pages} {481} (\bibinfo {year}
  {2012})}\BibitemShut {NoStop}%
\bibitem [{\citenamefont {Toyota}\ \emph {et~al.}(2011)\citenamefont {Toyota},
  \citenamefont {Head}, \citenamefont {Schmidt},\ and\ \citenamefont
  {Mizuno}}]{toyota2011}%
  \BibitemOpen
  \bibfield  {author} {\bibinfo {author} {\bibfnamefont {T.}~\bibnamefont
  {Toyota}}, \bibinfo {author} {\bibfnamefont {D.~A.}\ \bibnamefont {Head}},
  \bibinfo {author} {\bibfnamefont {C.~F.}\ \bibnamefont {Schmidt}}, \ and\
  \bibinfo {author} {\bibfnamefont {D.}~\bibnamefont {Mizuno}},\ }\href@noop {}
  {\bibfield  {journal} {\bibinfo  {journal} {Soft Matter}\ }\textbf {\bibinfo
  {volume} {7}},\ \bibinfo {pages} {3234} (\bibinfo {year} {2011})}\BibitemShut
  {NoStop}%
\bibitem [{\citenamefont {Yu}\ \emph {et~al.}(2013)\citenamefont {Yu},
  \citenamefont {Guan}, \citenamefont {Chen}, \citenamefont {Bae},\ and\
  \citenamefont {Granick}}]{yu2013}%
  \BibitemOpen
  \bibfield  {author} {\bibinfo {author} {\bibfnamefont {C.}~\bibnamefont
  {Yu}}, \bibinfo {author} {\bibfnamefont {J.}~\bibnamefont {Guan}}, \bibinfo
  {author} {\bibfnamefont {K.}~\bibnamefont {Chen}}, \bibinfo {author}
  {\bibfnamefont {S.~C.}\ \bibnamefont {Bae}}, \ and\ \bibinfo {author}
  {\bibfnamefont {S.}~\bibnamefont {Granick}},\ }\href@noop {} {\bibfield
  {journal} {\bibinfo  {journal} {ACS Nano}\ }\textbf {\bibinfo {volume} {7}},\
  \bibinfo {pages} {9735} (\bibinfo {year} {2013})}\BibitemShut {NoStop}%
\bibitem [{\citenamefont {Yu}\ and\ \citenamefont {Granick}(2014)}]{yu2014}%
  \BibitemOpen
  \bibfield  {author} {\bibinfo {author} {\bibfnamefont {C.}~\bibnamefont
  {Yu}}\ and\ \bibinfo {author} {\bibfnamefont {S.}~\bibnamefont {Granick}},\
  }\href@noop {} {\bibfield  {journal} {\bibinfo  {journal} {Langmuir}\
  }\textbf {\bibinfo {volume} {30}},\ \bibinfo {pages} {14538} (\bibinfo {year}
  {2014})}\BibitemShut {NoStop}%
\bibitem [{\citenamefont {Chakraborty}\ and\ \citenamefont
  {Roichman}(2020)}]{chakraborty2020}%
  \BibitemOpen
  \bibfield  {author} {\bibinfo {author} {\bibfnamefont {I.}~\bibnamefont
  {Chakraborty}}\ and\ \bibinfo {author} {\bibfnamefont {Y.}~\bibnamefont
  {Roichman}},\ }\href@noop {} {\bibfield  {journal} {\bibinfo  {journal}
  {Physical Review Research}\ }\textbf {\bibinfo {volume} {2}},\ \bibinfo
  {pages} {022020} (\bibinfo {year} {2020})}\BibitemShut {NoStop}%
\bibitem [{\citenamefont {Weeks}\ \emph {et~al.}(2000)\citenamefont {Weeks},
  \citenamefont {Crocker}, \citenamefont {Levitt}, \citenamefont {Schofield},\
  and\ \citenamefont {Weitz}}]{weeks2000}%
  \BibitemOpen
  \bibfield  {author} {\bibinfo {author} {\bibfnamefont {E.~R.}\ \bibnamefont
  {Weeks}}, \bibinfo {author} {\bibfnamefont {J.~C.}\ \bibnamefont {Crocker}},
  \bibinfo {author} {\bibfnamefont {A.~C.}\ \bibnamefont {Levitt}}, \bibinfo
  {author} {\bibfnamefont {A.}~\bibnamefont {Schofield}}, \ and\ \bibinfo
  {author} {\bibfnamefont {D.~A.}\ \bibnamefont {Weitz}},\ }\href@noop {}
  {\bibfield  {journal} {\bibinfo  {journal} {Science}\ }\textbf {\bibinfo
  {volume} {287}},\ \bibinfo {pages} {627} (\bibinfo {year}
  {2000})}\BibitemShut {NoStop}%
\bibitem [{\citenamefont {Wagner}\ \emph {et~al.}(2017)\citenamefont {Wagner},
  \citenamefont {Turner}, \citenamefont {Rubinstein}, \citenamefont
  {McKinley},\ and\ \citenamefont {Ribbeck}}]{wagner2017}%
  \BibitemOpen
  \bibfield  {author} {\bibinfo {author} {\bibfnamefont {C.~E.}\ \bibnamefont
  {Wagner}}, \bibinfo {author} {\bibfnamefont {B.~S.}\ \bibnamefont {Turner}},
  \bibinfo {author} {\bibfnamefont {M.}~\bibnamefont {Rubinstein}}, \bibinfo
  {author} {\bibfnamefont {G.~H.}\ \bibnamefont {McKinley}}, \ and\ \bibinfo
  {author} {\bibfnamefont {K.}~\bibnamefont {Ribbeck}},\ }\href@noop {}
  {\bibfield  {journal} {\bibinfo  {journal} {Biomacromolecules}\ }\textbf
  {\bibinfo {volume} {18}},\ \bibinfo {pages} {3654} (\bibinfo {year}
  {2017})}\BibitemShut {NoStop}%
\bibitem [{\citenamefont {Jeon}\ \emph {et~al.}(2016)\citenamefont {Jeon},
  \citenamefont {Javanainen}, \citenamefont {Martinez-Seara}, \citenamefont
  {Metzler},\ and\ \citenamefont {Vattulainen}}]{jeon2016}%
  \BibitemOpen
  \bibfield  {author} {\bibinfo {author} {\bibfnamefont {J.-H.}\ \bibnamefont
  {Jeon}}, \bibinfo {author} {\bibfnamefont {M.}~\bibnamefont {Javanainen}},
  \bibinfo {author} {\bibfnamefont {H.}~\bibnamefont {Martinez-Seara}},
  \bibinfo {author} {\bibfnamefont {R.}~\bibnamefont {Metzler}}, \ and\
  \bibinfo {author} {\bibfnamefont {I.}~\bibnamefont {Vattulainen}},\
  }\href@noop {} {\bibfield  {journal} {\bibinfo  {journal} {Physical Review
  X}\ }\textbf {\bibinfo {volume} {6}},\ \bibinfo {pages} {021006} (\bibinfo
  {year} {2016})}\BibitemShut {NoStop}%
\bibitem [{\citenamefont {Yamamoto}\ \emph {et~al.}(2017)\citenamefont
  {Yamamoto}, \citenamefont {Akimoto}, \citenamefont {Kalli}, \citenamefont
  {Yasuoka},\ and\ \citenamefont {Sansom}}]{yamamoto2017}%
  \BibitemOpen
  \bibfield  {author} {\bibinfo {author} {\bibfnamefont {E.}~\bibnamefont
  {Yamamoto}}, \bibinfo {author} {\bibfnamefont {T.}~\bibnamefont {Akimoto}},
  \bibinfo {author} {\bibfnamefont {A.~C.}\ \bibnamefont {Kalli}}, \bibinfo
  {author} {\bibfnamefont {K.}~\bibnamefont {Yasuoka}}, \ and\ \bibinfo
  {author} {\bibfnamefont {M.~S.}\ \bibnamefont {Sansom}},\ }\href@noop {}
  {\bibfield  {journal} {\bibinfo  {journal} {Science advances}\ }\textbf
  {\bibinfo {volume} {3}},\ \bibinfo {pages} {e1601871} (\bibinfo {year}
  {2017})}\BibitemShut {NoStop}%
\bibitem [{\citenamefont {Stylianidou}\ \emph {et~al.}(2014)\citenamefont
  {Stylianidou}, \citenamefont {Kuwada},\ and\ \citenamefont
  {Wiggins}}]{stylianidou2014}%
  \BibitemOpen
  \bibfield  {author} {\bibinfo {author} {\bibfnamefont {S.}~\bibnamefont
  {Stylianidou}}, \bibinfo {author} {\bibfnamefont {N.~J.}\ \bibnamefont
  {Kuwada}}, \ and\ \bibinfo {author} {\bibfnamefont {P.~A.}\ \bibnamefont
  {Wiggins}},\ }\href@noop {} {\bibfield  {journal} {\bibinfo  {journal}
  {Biophysical journal}\ }\textbf {\bibinfo {volume} {107}},\ \bibinfo {pages}
  {2684} (\bibinfo {year} {2014})}\BibitemShut {NoStop}%
\bibitem [{\citenamefont {Parry}\ \emph {et~al.}(2014)\citenamefont {Parry},
  \citenamefont {Surovtsev}, \citenamefont {Cabeen}, \citenamefont {O’Hern},
  \citenamefont {Dufresne},\ and\ \citenamefont {Jacobs-Wagner}}]{parry2014}%
  \BibitemOpen
  \bibfield  {author} {\bibinfo {author} {\bibfnamefont {B.~R.}\ \bibnamefont
  {Parry}}, \bibinfo {author} {\bibfnamefont {I.~V.}\ \bibnamefont
  {Surovtsev}}, \bibinfo {author} {\bibfnamefont {M.~T.}\ \bibnamefont
  {Cabeen}}, \bibinfo {author} {\bibfnamefont {C.~S.}\ \bibnamefont
  {O’Hern}}, \bibinfo {author} {\bibfnamefont {E.~R.}\ \bibnamefont
  {Dufresne}}, \ and\ \bibinfo {author} {\bibfnamefont {C.}~\bibnamefont
  {Jacobs-Wagner}},\ }\href@noop {} {\bibfield  {journal} {\bibinfo  {journal}
  {Cell}\ }\textbf {\bibinfo {volume} {156}},\ \bibinfo {pages} {183} (\bibinfo
  {year} {2014})}\BibitemShut {NoStop}%
\bibitem [{\citenamefont {Munder}\ \emph {et~al.}(2016)\citenamefont {Munder},
  \citenamefont {Midtvedt}, \citenamefont {Franzmann}, \citenamefont {Nuske},
  \citenamefont {Otto}, \citenamefont {Herbig}, \citenamefont {Ulbricht},
  \citenamefont {M{\"u}ller}, \citenamefont {Taubenberger}, \citenamefont
  {Maharana} \emph {et~al.}}]{munder2016}%
  \BibitemOpen
  \bibfield  {author} {\bibinfo {author} {\bibfnamefont {M.~C.}\ \bibnamefont
  {Munder}}, \bibinfo {author} {\bibfnamefont {D.}~\bibnamefont {Midtvedt}},
  \bibinfo {author} {\bibfnamefont {T.}~\bibnamefont {Franzmann}}, \bibinfo
  {author} {\bibfnamefont {E.}~\bibnamefont {Nuske}}, \bibinfo {author}
  {\bibfnamefont {O.}~\bibnamefont {Otto}}, \bibinfo {author} {\bibfnamefont
  {M.}~\bibnamefont {Herbig}}, \bibinfo {author} {\bibfnamefont
  {E.}~\bibnamefont {Ulbricht}}, \bibinfo {author} {\bibfnamefont
  {P.}~\bibnamefont {M{\"u}ller}}, \bibinfo {author} {\bibfnamefont
  {A.}~\bibnamefont {Taubenberger}}, \bibinfo {author} {\bibfnamefont
  {S.}~\bibnamefont {Maharana}},  \emph {et~al.},\ }\href@noop {} {\bibfield
  {journal} {\bibinfo  {journal} {elife}\ }\textbf {\bibinfo {volume} {5}},\
  \bibinfo {pages} {e09347} (\bibinfo {year} {2016})}\BibitemShut {NoStop}%
\bibitem [{\citenamefont {Cherstvy}\ \emph {et~al.}(2018)\citenamefont
  {Cherstvy}, \citenamefont {Nagel}, \citenamefont {Beta},\ and\ \citenamefont
  {Metzler}}]{cherstvy2018}%
  \BibitemOpen
  \bibfield  {author} {\bibinfo {author} {\bibfnamefont {A.~G.}\ \bibnamefont
  {Cherstvy}}, \bibinfo {author} {\bibfnamefont {O.}~\bibnamefont {Nagel}},
  \bibinfo {author} {\bibfnamefont {C.}~\bibnamefont {Beta}}, \ and\ \bibinfo
  {author} {\bibfnamefont {R.}~\bibnamefont {Metzler}},\ }\href@noop {}
  {\bibfield  {journal} {\bibinfo  {journal} {Physical Chemistry Chemical
  Physics}\ }\textbf {\bibinfo {volume} {20}},\ \bibinfo {pages} {23034}
  (\bibinfo {year} {2018})}\BibitemShut {NoStop}%
\bibitem [{\citenamefont {Li}\ \emph {et~al.}(2019)\citenamefont {Li},
  \citenamefont {Marchesoni}, \citenamefont {Debnath},\ and\ \citenamefont
  {Ghosh}}]{li2019}%
  \BibitemOpen
  \bibfield  {author} {\bibinfo {author} {\bibfnamefont {Y.}~\bibnamefont
  {Li}}, \bibinfo {author} {\bibfnamefont {F.}~\bibnamefont {Marchesoni}},
  \bibinfo {author} {\bibfnamefont {D.}~\bibnamefont {Debnath}}, \ and\
  \bibinfo {author} {\bibfnamefont {P.~K.}\ \bibnamefont {Ghosh}},\ }\href@noop
  {} {\bibfield  {journal} {\bibinfo  {journal} {Physical Review Research}\
  }\textbf {\bibinfo {volume} {1}},\ \bibinfo {pages} {033003} (\bibinfo {year}
  {2019})}\BibitemShut {NoStop}%
\bibitem [{\citenamefont {Cuetos}\ \emph {et~al.}(2018)\citenamefont {Cuetos},
  \citenamefont {Morillo},\ and\ \citenamefont {Patti}}]{cuetos2018}%
  \BibitemOpen
  \bibfield  {author} {\bibinfo {author} {\bibfnamefont {A.}~\bibnamefont
  {Cuetos}}, \bibinfo {author} {\bibfnamefont {N.}~\bibnamefont {Morillo}}, \
  and\ \bibinfo {author} {\bibfnamefont {A.}~\bibnamefont {Patti}},\
  }\href@noop {} {\bibfield  {journal} {\bibinfo  {journal} {Physical Review
  E}\ }\textbf {\bibinfo {volume} {98}},\ \bibinfo {pages} {042129} (\bibinfo
  {year} {2018})}\BibitemShut {NoStop}%
\bibitem [{\citenamefont {Hapca}\ \emph {et~al.}(2008)\citenamefont {Hapca},
  \citenamefont {Crawford},\ and\ \citenamefont {Young}}]{hapca2008}%
  \BibitemOpen
  \bibfield  {author} {\bibinfo {author} {\bibfnamefont {S.}~\bibnamefont
  {Hapca}}, \bibinfo {author} {\bibfnamefont {J.~W.}\ \bibnamefont {Crawford}},
  \ and\ \bibinfo {author} {\bibfnamefont {I.~M.}\ \bibnamefont {Young}},\
  }\href@noop {} {\bibfield  {journal} {\bibinfo  {journal} {Journal of the
  Royal Society Interface}\ }\textbf {\bibinfo {volume} {6}},\ \bibinfo {pages}
  {111} (\bibinfo {year} {2009})}\BibitemShut {NoStop}%
\bibitem [{\citenamefont {Pastore}\ \emph {et~al.}(2021)\citenamefont
  {Pastore}, \citenamefont {Ciarlo}, \citenamefont {Pesce}, \citenamefont
  {Greco},\ and\ \citenamefont {Sasso}}]{pastore2021rapid}%
  \BibitemOpen
  \bibfield  {author} {\bibinfo {author} {\bibfnamefont {R.}~\bibnamefont
  {Pastore}}, \bibinfo {author} {\bibfnamefont {A.}~\bibnamefont {Ciarlo}},
  \bibinfo {author} {\bibfnamefont {G.}~\bibnamefont {Pesce}}, \bibinfo
  {author} {\bibfnamefont {F.}~\bibnamefont {Greco}}, \ and\ \bibinfo {author}
  {\bibfnamefont {A.}~\bibnamefont {Sasso}},\ }\href@noop {} {\bibfield
  {journal} {\bibinfo  {journal} {Physical Review Letters}\ }\textbf {\bibinfo
  {volume} {126}},\ \bibinfo {pages} {158003} (\bibinfo {year}
  {2021})}\BibitemShut {NoStop}%
\bibitem [{\citenamefont {Pastore}\ and\ \citenamefont
  {Raos}(2015)}]{pastore2015}%
  \BibitemOpen
  \bibfield  {author} {\bibinfo {author} {\bibfnamefont {R.}~\bibnamefont
  {Pastore}}\ and\ \bibinfo {author} {\bibfnamefont {G.}~\bibnamefont {Raos}},\
  }\href@noop {} {\bibfield  {journal} {\bibinfo  {journal} {Soft Matter}\
  }\textbf {\bibinfo {volume} {11}},\ \bibinfo {pages} {8083} (\bibinfo {year}
  {2015})}\BibitemShut {NoStop}%
\bibitem [{\citenamefont {Miotto}\ \emph {et~al.}(2021)\citenamefont {Miotto},
  \citenamefont {Pigolotti}, \citenamefont {Chechkin},\ and\ \citenamefont
  {Rold{\'a}n-Vargas}}]{miotto2021length}%
  \BibitemOpen
  \bibfield  {author} {\bibinfo {author} {\bibfnamefont {J.~M.}\ \bibnamefont
  {Miotto}}, \bibinfo {author} {\bibfnamefont {S.}~\bibnamefont {Pigolotti}},
  \bibinfo {author} {\bibfnamefont {A.~V.}\ \bibnamefont {Chechkin}}, \ and\
  \bibinfo {author} {\bibfnamefont {S.}~\bibnamefont {Rold{\'a}n-Vargas}},\
  }\href@noop {} {\bibfield  {journal} {\bibinfo  {journal} {Physical Review
  X}\ }\textbf {\bibinfo {volume} {11}},\ \bibinfo {pages} {031002} (\bibinfo
  {year} {2021})}\BibitemShut {NoStop}%
\bibitem [{\citenamefont {Rusciano}\ \emph {et~al.}(2022)\citenamefont
  {Rusciano}, \citenamefont {Pastore},\ and\ \citenamefont
  {Greco}}]{pastore2022}%
  \BibitemOpen
  \bibfield  {author} {\bibinfo {author} {\bibfnamefont {F.}~\bibnamefont
  {Rusciano}}, \bibinfo {author} {\bibfnamefont {R.}~\bibnamefont {Pastore}}, \
  and\ \bibinfo {author} {\bibfnamefont {F.}~\bibnamefont {Greco}},\
  }\href@noop {} {\bibfield  {journal} {\bibinfo  {journal} {Physical Review
  Letters}\ }\textbf {\bibinfo {volume} {128}},\ \bibinfo {pages} {168001}
  (\bibinfo {year} {2022})}\BibitemShut {NoStop}%
\bibitem [{\citenamefont {Lanoisel\'ee}\ \emph {et~al.}(2018)\citenamefont
  {Lanoisel\'ee}, \citenamefont {Moutal},\ and\ \citenamefont
  {Grebenkov}}]{grebenkov2018}%
  \BibitemOpen
  \bibfield  {author} {\bibinfo {author} {\bibfnamefont {Y.}~\bibnamefont
  {Lanoisel\'ee}}, \bibinfo {author} {\bibfnamefont {N.}~\bibnamefont
  {Moutal}}, \ and\ \bibinfo {author} {\bibfnamefont {D.~S.}\ \bibnamefont
  {Grebenkov}},\ }\href@noop {} {\bibfield  {journal} {\bibinfo  {journal}
  {Nat. Comm.}\ }\textbf {\bibinfo {volume} {9}},\ \bibinfo {pages} {4398}
  (\bibinfo {year} {2018})}\BibitemShut {NoStop}%
\bibitem [{\citenamefont {Grebenkov}(2019)}]{grebenkov2019}%
  \BibitemOpen
  \bibfield  {author} {\bibinfo {author} {\bibfnamefont {D.~S.}\ \bibnamefont
  {Grebenkov}},\ }\href@noop {} {\bibfield  {journal} {\bibinfo  {journal} {J.
  Phys. A}\ }\textbf {\bibinfo {volume} {52}},\ \bibinfo {pages} {174001}
  (\bibinfo {year} {2019})}\BibitemShut {NoStop}%
\bibitem [{\citenamefont {Sposini}\ \emph {et~al.}(2019)\citenamefont
  {Sposini}, \citenamefont {Chechkin},\ and\ \citenamefont
  {Metzler}}]{sposini2019first}%
  \BibitemOpen
  \bibfield  {author} {\bibinfo {author} {\bibfnamefont {V.}~\bibnamefont
  {Sposini}}, \bibinfo {author} {\bibfnamefont {A.}~\bibnamefont {Chechkin}}, \
  and\ \bibinfo {author} {\bibfnamefont {R.}~\bibnamefont {Metzler}},\
  }\href@noop {} {\bibfield  {journal} {\bibinfo  {journal} {Journal of Physics
  A: Mathematical and Theoretical}\ }\textbf {\bibinfo {volume} {52}},\
  \bibinfo {pages} {04LT01} (\bibinfo {year} {2019})}\BibitemShut {NoStop}%
\bibitem [{\citenamefont {Chechkin}\ \emph {et~al.}(2017)\citenamefont
  {Chechkin}, \citenamefont {Seno}, \citenamefont {Metzler},\ and\
  \citenamefont {Sokolov}}]{chechkin2017}%
  \BibitemOpen
  \bibfield  {author} {\bibinfo {author} {\bibfnamefont {A.~V.}\ \bibnamefont
  {Chechkin}}, \bibinfo {author} {\bibfnamefont {F.}~\bibnamefont {Seno}},
  \bibinfo {author} {\bibfnamefont {R.}~\bibnamefont {Metzler}}, \ and\
  \bibinfo {author} {\bibfnamefont {I.~M.}\ \bibnamefont {Sokolov}},\
  }\href@noop {} {\bibfield  {journal} {\bibinfo  {journal} {Physical Review
  X}\ }\textbf {\bibinfo {volume} {7}},\ \bibinfo {pages} {021002} (\bibinfo
  {year} {2017})}\BibitemShut {NoStop}%
\bibitem [{\citenamefont {Nampoothiri}\ \emph {et~al.}(2021)\citenamefont
  {Nampoothiri}, \citenamefont {Orlandini}, \citenamefont {Seno},\ and\
  \citenamefont {Baldovin}}]{nampoothiri2021}%
  \BibitemOpen
  \bibfield  {author} {\bibinfo {author} {\bibfnamefont {S.}~\bibnamefont
  {Nampoothiri}}, \bibinfo {author} {\bibfnamefont {E.}~\bibnamefont
  {Orlandini}}, \bibinfo {author} {\bibfnamefont {F.}~\bibnamefont {Seno}}, \
  and\ \bibinfo {author} {\bibfnamefont {F.}~\bibnamefont {Baldovin}},\
  }\href@noop {} {\bibfield  {journal} {\bibinfo  {journal} {Physical Review
  E}\ }\textbf {\bibinfo {volume} {104}},\ \bibinfo {pages} {L062501} (\bibinfo
  {year} {2021})}\BibitemShut {NoStop}%
\bibitem [{\citenamefont {Nampoothiri}\ \emph {et~al.}(2022)\citenamefont
  {Nampoothiri}, \citenamefont {Orlandini}, \citenamefont {Seno},\ and\
  \citenamefont {Baldovin}}]{nampoothiri2022}%
  \BibitemOpen
  \bibfield  {author} {\bibinfo {author} {\bibfnamefont {S.}~\bibnamefont
  {Nampoothiri}}, \bibinfo {author} {\bibfnamefont {E.}~\bibnamefont
  {Orlandini}}, \bibinfo {author} {\bibfnamefont {F.}~\bibnamefont {Seno}}, \
  and\ \bibinfo {author} {\bibfnamefont {F.}~\bibnamefont {Baldovin}},\
  }\href@noop {} {\bibfield  {journal} {\bibinfo  {journal} {New J. Phys.}\
  }\textbf {\bibinfo {volume} {24}},\ \bibinfo {pages} {023003} (\bibinfo
  {year} {2022})}\BibitemShut {NoStop}%
\bibitem [{\citenamefont {Marcone}\ \emph {et~al.}(2022)\citenamefont
  {Marcone}, \citenamefont {Nampoothiri}, \citenamefont {Orlandini},
  \citenamefont {Seno},\ and\ \citenamefont {Baldovin}}]{marcone2022}%
  \BibitemOpen
  \bibfield  {author} {\bibinfo {author} {\bibfnamefont {B.}~\bibnamefont
  {Marcone}}, \bibinfo {author} {\bibfnamefont {S.}~\bibnamefont
  {Nampoothiri}}, \bibinfo {author} {\bibfnamefont {E.}~\bibnamefont
  {Orlandini}}, \bibinfo {author} {\bibfnamefont {F.}~\bibnamefont {Seno}}, \
  and\ \bibinfo {author} {\bibfnamefont {F.}~\bibnamefont {Baldovin}},\
  }\href@noop {} {\bibfield  {journal} {\bibinfo  {journal} {J. Phys. A: Math.
  Theor.}\ }\textbf {\bibinfo {volume} {55}},\ \bibinfo {pages} {354003}
  (\bibinfo {year} {2022})}\BibitemShut {NoStop}%
\bibitem [{\citenamefont {Flory}(1953)}]{flory1953}%
  \BibitemOpen
  \bibfield  {author} {\bibinfo {author} {\bibfnamefont {P.}~\bibnamefont
  {Flory}},\ }\href@noop {} {\emph {\bibinfo {title} {Principles of Polymer
  Chemistry}}}\ (\bibinfo  {publisher} {Cornell University Press},\ \bibinfo
  {year} {1953})\BibitemShut {NoStop}%
\bibitem [{\citenamefont {Cosgrove}(2005)}]{cosgrove2005}%
  \BibitemOpen
  \bibfield  {author} {\bibinfo {author} {\bibfnamefont {T.}~\bibnamefont
  {Cosgrove}},\ }\href@noop {} {\emph {\bibinfo {title} {Colloid Science
  Principles, Methods and Applications}}}\ (\bibinfo  {publisher} {Blackwell
  Publishing, Oxford, UK},\ \bibinfo {year} {2005})\BibitemShut {NoStop}%
\bibitem [{\citenamefont {Odian}(2004)}]{odian2004}%
  \BibitemOpen
  \bibfield  {author} {\bibinfo {author} {\bibfnamefont {G.}~\bibnamefont
  {Odian}},\ }\href@noop {} {\emph {\bibinfo {title} {Principles of
  Polymerization}}}\ (\bibinfo  {publisher} {John Wiley \& Sons},\ \bibinfo
  {year} {2004})\BibitemShut {NoStop}%
\bibitem [{\citenamefont {Klafter}\ and\ \citenamefont
  {Sokolov}(2011)}]{klafter2011}%
  \BibitemOpen
  \bibfield  {author} {\bibinfo {author} {\bibfnamefont {J.}~\bibnamefont
  {Klafter}}\ and\ \bibinfo {author} {\bibfnamefont {I.}~\bibnamefont
  {Sokolov}},\ }\href@noop {} {\emph {\bibinfo {title} {First Steps in Random
  Walks: From Tools to Applications}}}\ (\bibinfo  {publisher} {Oxford
  University Press},\ \bibinfo {year} {2011})\BibitemShut {NoStop}%
\bibitem [{\citenamefont {Barkai}\ and\ \citenamefont
  {Burov}(2020)}]{barkai2020}%
  \BibitemOpen
  \bibfield  {author} {\bibinfo {author} {\bibfnamefont {E.}~\bibnamefont
  {Barkai}}\ and\ \bibinfo {author} {\bibfnamefont {S.}~\bibnamefont {Burov}},\
  }\href@noop {} {\bibfield  {journal} {\bibinfo  {journal} {Physical Review
  Letters}\ }\textbf {\bibinfo {volume} {124}},\ \bibinfo {pages} {060603}
  (\bibinfo {year} {2020})}\BibitemShut {NoStop}%
\bibitem [{\citenamefont {Wang}\ \emph {et~al.}(2020)\citenamefont {Wang},
  \citenamefont {Barkai},\ and\ \citenamefont {Burov}}]{wang2020}%
  \BibitemOpen
  \bibfield  {author} {\bibinfo {author} {\bibfnamefont {W.}~\bibnamefont
  {Wang}}, \bibinfo {author} {\bibfnamefont {E.}~\bibnamefont {Barkai}}, \ and\
  \bibinfo {author} {\bibfnamefont {S.}~\bibnamefont {Burov}},\ }\href@noop {}
  {\bibfield  {journal} {\bibinfo  {journal} {Entropy}\ }\textbf {\bibinfo
  {volume} {22}},\ \bibinfo {pages} {697} (\bibinfo {year} {2020})}\BibitemShut
  {NoStop}%
\bibitem [{\citenamefont {Pacheco-Pozo}\ and\ \citenamefont
  {Sokolov}(2021{\natexlab{a}})}]{sokolov2021}%
  \BibitemOpen
  \bibfield  {author} {\bibinfo {author} {\bibfnamefont {A.}~\bibnamefont
  {Pacheco-Pozo}}\ and\ \bibinfo {author} {\bibfnamefont {I.~M.}\ \bibnamefont
  {Sokolov}},\ }\href@noop {} {\bibfield  {journal} {\bibinfo  {journal}
  {Physical Review Letters}\ }\textbf {\bibinfo {volume} {127}},\ \bibinfo
  {pages} {120601} (\bibinfo {year} {2021}{\natexlab{a}})}\BibitemShut
  {NoStop}%
\bibitem [{\citenamefont {Pacheco-Pozo}\ and\ \citenamefont
  {Sokolov}(2021{\natexlab{b}})}]{sokolov2021ld}%
  \BibitemOpen
  \bibfield  {author} {\bibinfo {author} {\bibfnamefont {A.}~\bibnamefont
  {Pacheco-Pozo}}\ and\ \bibinfo {author} {\bibfnamefont {I.~M.}\ \bibnamefont
  {Sokolov}},\ }\href@noop {} {\bibfield  {journal} {\bibinfo  {journal}
  {Physical Review E}\ }\textbf {\bibinfo {volume} {103}},\ \bibinfo {pages}
  {042116} (\bibinfo {year} {2021}{\natexlab{b}})}\BibitemShut {NoStop}%
\bibitem [{\citenamefont {Rudin}(1987)}]{rudin1987}%
  \BibitemOpen
  \bibfield  {author} {\bibinfo {author} {\bibfnamefont {W.}~\bibnamefont
  {Rudin}},\ }\href@noop {} {\emph {\bibinfo {title} {Real and complex
  analysis}}}\ (\bibinfo  {publisher} {McGraw-Hill},\ \bibinfo {year}
  {1987})\BibitemShut {NoStop}%
\bibitem [{\citenamefont {Sposini}\ \emph {et~al.}(2023)\citenamefont
  {Sposini}, \citenamefont {Nampoothiri}, \citenamefont {Chechkin},
  \citenamefont {Orlandini}, \citenamefont {Seno},\ and\ \citenamefont
  {Baldovin}}]{sposini2023prl}%
  \BibitemOpen
  \bibfield  {author} {\bibinfo {author} {\bibfnamefont {V.}~\bibnamefont
  {Sposini}}, \bibinfo {author} {\bibfnamefont {S.}~\bibnamefont
  {Nampoothiri}}, \bibinfo {author} {\bibfnamefont {A.}~\bibnamefont
  {Chechkin}}, \bibinfo {author} {\bibfnamefont {E.}~\bibnamefont {Orlandini}},
  \bibinfo {author} {\bibfnamefont {F.}~\bibnamefont {Seno}}, \ and\ \bibinfo
  {author} {\bibfnamefont {F.}~\bibnamefont {Baldovin}},\ }\href@noop {}
  {\bibfield  {journal} {\bibinfo  {journal} {Physical Review Letters}\
  }\textbf {\bibinfo {volume} {132}}, \ \bibinfo {pages}
  {117101} (\bibinfo {year}
  {2024})}\BibitemShut {NoStop}%
\bibitem [{\citenamefont {Feller}(1968)}]{Feller1968}%
  \BibitemOpen
  \bibfield  {author} {\bibinfo {author} {\bibfnamefont {W.}~\bibnamefont
  {Feller}},\ }\href@noop {} {\emph {\bibinfo {title} {An Introduction to
  Probability Theory and Its Applications}}}\ (\bibinfo  {publisher} {John
  Wiley \& Sons},\ \bibinfo {year} {1968})\BibitemShut {NoStop}%
\bibitem [{\citenamefont {Bochner}(2020)}]{bochner2020harmonic}%
  \BibitemOpen
  \bibfield  {author} {\bibinfo {author} {\bibfnamefont {S.}~\bibnamefont
  {Bochner}},\ }\href@noop {} {\emph {\bibinfo {title} {Harmonic analysis and
  the theory of probability}}}\ (\bibinfo  {publisher} {University of
  California press},\ \bibinfo {year} {2020})\BibitemShut {NoStop}%
\bibitem [{\citenamefont {Heston}(1993)}]{heston1993}%
  \BibitemOpen
  \bibfield  {author} {\bibinfo {author} {\bibfnamefont {S.}~\bibnamefont
  {Heston}},\ }\href@noop {} {\bibfield  {journal} {\bibinfo  {journal} {Rev.
  Financial Studies}\ }\textbf {\bibinfo {volume} {6}},\ \bibinfo {pages} {327}
  (\bibinfo {year} {1993})}\BibitemShut {NoStop}%
\bibitem [{\citenamefont {Fouqu\'e}\ \emph {et~al.}(2000)\citenamefont
  {Fouqu\'e}, \citenamefont {Papanicolaou},\ and\ \citenamefont
  {Sircar}}]{sircar2000}%
  \BibitemOpen
  \bibfield  {author} {\bibinfo {author} {\bibfnamefont {J.-P.}\ \bibnamefont
  {Fouqu\'e}}, \bibinfo {author} {\bibfnamefont {G.}~\bibnamefont
  {Papanicolaou}}, \ and\ \bibinfo {author} {\bibfnamefont {K.}~\bibnamefont
  {Sircar}},\ }\href@noop {} {\emph {\bibinfo {title} {Derivatives in Financial
  Markets with Stochastic Volatility}}}\ (\bibinfo  {publisher} {Cambridge
  University Press, Cambridge, England},\ \bibinfo {year} {2000})\BibitemShut
  {NoStop}%
\bibitem [{\citenamefont {de~Gennes}(1979)}]{deGennes1979}%
  \BibitemOpen
  \bibfield  {author} {\bibinfo {author} {\bibfnamefont {P.-G.}\ \bibnamefont
  {de~Gennes}},\ }\href@noop {} {\emph {\bibinfo {title} {Scaling Concepts in
  Polymer Physics}}}\ (\bibinfo  {publisher} {Cornell University Press},\
  \bibinfo {year} {1979})\BibitemShut {NoStop}%
\bibitem [{\citenamefont {Doi}\ and\ \citenamefont {F}(1992)}]{Doi1992}%
  \BibitemOpen
  \bibfield  {author} {\bibinfo {author} {\bibfnamefont {M.}~\bibnamefont
  {Doi}}\ and\ \bibinfo {author} {\bibfnamefont {E.~S.}\ \bibnamefont {F}},\
  }\href@noop {} {\emph {\bibinfo {title} {The Theory of Polymer Dynamics}}}\
  (\bibinfo  {publisher} {Oxford University Press},\ \bibinfo {year}
  {1992})\BibitemShut {NoStop}%
\bibitem [{\citenamefont {Gillespie}(1977)}]{gillespie1977}%
  \BibitemOpen
  \bibfield  {author} {\bibinfo {author} {\bibfnamefont {D.~T.}\ \bibnamefont
  {Gillespie}},\ }\href@noop {} {\bibfield  {journal} {\bibinfo  {journal}
  {Journal of Physical Chemistry}\ }\textbf {\bibinfo {volume} {81}},\ \bibinfo
  {pages} {2340–2361} (\bibinfo {year} {1977})}\BibitemShut {NoStop}%
\bibitem [{\citenamefont {Beck}\ and\ \citenamefont {Cohen}(2003)}]{beck2003}%
  \BibitemOpen
  \bibfield  {author} {\bibinfo {author} {\bibfnamefont {C.}~\bibnamefont
  {Beck}}\ and\ \bibinfo {author} {\bibfnamefont {E.~G.}\ \bibnamefont
  {Cohen}},\ }\href@noop {} {\bibfield  {journal} {\bibinfo  {journal} {Physica
  A: Statistical mechanics and its applications}\ }\textbf {\bibinfo {volume}
  {322}},\ \bibinfo {pages} {267} (\bibinfo {year} {2003})}\BibitemShut
  {NoStop}%
\bibitem [{\citenamefont {Beck}(2006)}]{beck2006}%
  \BibitemOpen
  \bibfield  {author} {\bibinfo {author} {\bibfnamefont {C.}~\bibnamefont
  {Beck}},\ }\href@noop {} {\bibfield  {journal} {\bibinfo  {journal} {Progress
  of Theoretical Physics Supplement}\ }\textbf {\bibinfo {volume} {162}},\
  \bibinfo {pages} {29} (\bibinfo {year} {2006})}\BibitemShut {NoStop}%
\bibitem [{\citenamefont {Touchette}(2009)}]{touchette2009}%
  \BibitemOpen
  \bibfield  {author} {\bibinfo {author} {\bibfnamefont {H.}~\bibnamefont
  {Touchette}},\ }\href@noop {} {\bibfield  {journal} {\bibinfo  {journal}
  {Physics Reports}\ }\textbf {\bibinfo {volume} {478}},\ \bibinfo {pages} {1}
  (\bibinfo {year} {2009})}\BibitemShut {NoStop}%
\bibitem [{\citenamefont {Redner}(2001)}]{redner2001}%
  \BibitemOpen
  \bibfield  {author} {\bibinfo {author} {\bibfnamefont {S.}~\bibnamefont
  {Redner}},\ }\href@noop {} {\emph {\bibinfo {title} {A guide to first passage
  processes}}}\ (\bibinfo  {publisher} {Cambridge University Press},\ \bibinfo
  {year} {2001})\BibitemShut {NoStop}%
\end{thebibliography}
\end{document}